\documentclass[eqsecnum,showpacs,showkeys,prd]{re
vtex4}
\usepackage{amsmath}
\usepackage[dvips]{graphicx}
\usepackage{color}
\begin{document}
		  
\title{Accelerating Observers, Area and Entropy}

\author{Jarmo M\"akel\"a} 
\email[Electronic address: 
]{jarmo.makela@phys.jyu.fi}  
\affiliation{Department of Physics, University of 
Jyv\"askyl\"a, PB 35 (YFL), FIN-40351 
Jyv\"askyl\"a, Finland}

\begin{abstract}
We consider an explicit example of a process, 
where the entropy carried by radiation through an 
accelerating spacelike two-plane is proportional 
to the decrease in the area of that two-plane 
even when the two-plane is not a part of any 
horizon of spacetime. Our results seem to support 
the view that entropy proportional to area is 
possessed not only by horizons but by all 
spacelike two-surfaces of spacetime.  
\end{abstract}


\maketitle
		  
\section{Introduction}
		  
		  A very interesting discovery was 
published by
                  Jacobson in 1995 \cite{eka}. He 
had found that, in a certain sense, Einstein's 
field equation may be viewed as a thermodynamic 
equation of state of spacetime and matter fields. 
This discovery was based on the fact that when 
the Rindler horizon of an accelerating observer 
emits the so called Unruh radiation, the area of 
the considered part of the Rindler horizon 
shrinks. Jacobson observed that if one {\it 
assumes} that the entropy carried away from the 
considered part of the Rindler horizon is, in 
natural units, exactly one quarter of the 
decrease in the area of that part, then 
Einstein's field equation follows from the first 
law of thermodynamics. In other words, it turned 
out that Einstein's field equation is just a 
consequence from the first law of thermodynamics, 
and an assumption that entropy equal to one 
quarter of the horizon area is possessed not only 
by black hole and cosmological horizons, but also 
by the Rindler horizon. This conclusion supports 
the view that classical general relativity 
actually describes
the thermodynamic properties of spacetime 
\cite{toka}. If one wants to construct a 
microscopic theory of gravitation, the correct 
way to do this may not be to quantize general 
relativity but to postulate an existence of 
certain microscopic constituents of spacetime 
obeying laws which, in the thermodynamic limit, 
reproduce Einstein's field equations. 

        An essential feature in Jacobson's 
analysis was that a {\it horizon} was looked at: 
When radiation is emitted by a horizon, the 
horizon shrinks, and the entropy of the horizon 
is converted to the entropy of the radiation. It 
is interesting to investigate, whether entropy 
would be possessed not only by horizons but by 
{\it all} spacelike two-surfaces of spacetime. 
Arguments supporting this claim have been given, 
for instance, in \cite{kolmas}, where it was 
proposed that every spacelike two-surface of 
spacetime possesses entropy which, in natural 
units, is one quarter of the area of that 
two-surface. If true, this proposal brings great 
clarity to the concept of gravitational entropy: 
Entropy is a property of spacetime, and the 
entropy of a spacelike two-surface is always the 
same, no matter whether that two-surface is a 
part of a horizon or not.

      To support the view that entropy is 
possessed not only by the horizons, but by all 
spacelike two-surfaces of spacetime we consider, 
in this paper, a slight modification of 
Jacobson's analysis. More precisely, we consider 
a plane parallel to the $xy$-plane, which is 
accelerating with a constant proper acceleration 
along the positive $z$-axis. We assume that, from 
the point of view an observer at rest with 
respect to the flat Minkowski coordinates $x$, 
$y$ and $z$, spacetime is filled with 
electromagnetic radiation in a thermal 
equilibrium, from which it follows that there is 
a certain net flow of radiation through the 
accelerating plane to the negative $z$-direction. 
We investigate the properties of this radiation, 
and it turns out that when radiation propagates 
through the accelerating plane, the plane 
shrinks. Using the first law of thermodynamics we 
shall arrive at an interesting conclusion that 
the entropy carried by Unruh radiation through 
the accelerating plane is exactly one half of the 
decrease in the area of that plane. In other 
words, it seems that we a have found a process 
where the entropy of a two-plane is converted to 
the entropy of radiation even when that two-plane 
is not a part of any horizon of spacetime. This 
result strongly supports the proposal that not 
only horizons, but all spacelike two-surfaces of 
spacetime possess a certain amount of entropy.

    This paper is organized as follows: In 
Section 2 we consider the energy flow and 
intensity of radiation through an accelerating 
plane. In Section 3 we investigate the change in 
the area of the accelerating two-plane when 
radiation flows through that plane. Finally, in 
Section 4 we bring to the stage the first law of 
thermodynamics. The first law of thermodynamics, 
together with the results of Sections 2 and 3, 
implies the desired relationship between area and 
entropy. We end our discussion with concluding 
remarks. Unless otherwise stated, we shall always 
use the natural units, where $\hbar=c=G=k_B=1$. 

\maketitle

\section{Flow of Energy}

      When flat Minkowski spacetime is filled 
with electromagnetic radiation in thermal 
equilibrium, there is no net flow of energy in 
any direction from the point of view of an 
observer at rest with respect to the flat 
Minkowski coordinates $t$, $x$, $y$, and $z$. The 
only non-zero components of the energy momentum 
stress tensor $T^{\mu\nu}$ of radiation are:
\begin{equation}
T^{00}=\rho,
\end{equation}
where $\rho$ is the energy density of radiation, 
and
\begin{equation}
T^{11}=T^{22}=T^{33}=p,
\end{equation}
where $p$ is the pressure of the radiation. 
Because for electromagnetic radiation
\begin{equation}
p=\frac{1}{3}\rho,
\end{equation}
we find that
\begin{equation}
T:=T^0_0+T^1_1+T^2_2+T^3_3=0.
\end{equation}
In other words, the tensor $T^{\mu\nu}$ is 
trace-free. This result will turn out very useful 
in our investigations.

       Although there is no net flow of energy 
from the point of view of an observer at rest 
with respect to the coordinates $x$, $y$ and $z$, 
there is, however, a net flow of energy to the 
negative $z$-direction from the point of view of 
an observer in a uniformly accelerating motion 
along the positive $z$-axis. The world line of 
such an observer satisfies the equation 
\begin{equation}
z^2-t^2=\frac{1}{a^2},
\end{equation}
where $a$ is the proper acceleration of the 
observer. This world line may be parametrized by 
the observer's proper time $\tau$ such that 
\begin{subequations}
\begin{eqnarray}
t&=&\frac{1}{a}\sinh(a\tau),  \\
z&=&\frac{1}{a}\cosh(a\tau). 
\end{eqnarray}
\end{subequations}
The time- and the $z$-coordinates of the 
accelerating observer are $t'$ and $z'$, 
respectively, such that for infinitesimal changes 
of $t'$ and $z'$:
\begin{subequations}
\begin{eqnarray}
dt'=d\tau &=&\frac{dt-v\,dz}{\sqrt{1-v^2}}, \\
dz' &=& \frac{dz-v\,dt}{\sqrt{1-v^2}}, 
\end{eqnarray}
\end{subequations}
where
\begin{equation}
v:=\tanh(a\tau).
\end{equation}
is the speed of the accelerating observer with 
respect to the observer at rest. The flow of 
energy, or intensity of radiation to the negative 
$z$-direction through the $xy$-plane at rest with 
respect to the accelerating observer is
\begin{equation}
I=-T'^{03}=-\frac{\partial x'^0}{\partial 
x^\mu}\frac{\partial x'^3}{\partial 
x^\nu}T^{\mu\nu}=\frac{1}{2}\sinh(2a\tau)(T^{00}+
T^{33}),
\end{equation}
and it follows from Eqs. (2.1)-(2.3) that
\begin{equation}
I=\frac{2}{3}\sinh(2a\tau)\rho.
\end{equation}
Hence we see that if we pick up a region with 
area $A$ from the $xy$-plane, the energy carried 
by radiation through that region during an 
infinitesimal proper time interval $d\tau$ is 
\begin{equation}
\delta Q = \frac{2}{3}A\rho\sinh(2a\tau)\,d\tau.
\end{equation}

\section{Change of Area}

In the previous Section we assumed that spacetime 
is flat, and filled with electromagnetic 
radiation in thermal equlibrium. Interaction 
between radiation and the geometry of spacetime, 
however, makes spacetime curved, and as a result 
there is a change in the area of the accelerating 
plane we considered in the previous Section. When 
we evaluate the change in the area of an 
accelerating plane in the course of time, we 
might use, of course, the so called Raychaudhuri 
equation \cite{neljas}, and that was indeed the equation on 
which Jacobson's investigations were based. 
Raychaudhuri equation, however, describes the 
behavior of {\it geodesics} of spacetime, and 
because an accelerating observer does not move 
along a spacetime geodesic, it is not quite clear 
how to apply that equation. Because of the 
conceptual problems involved in the Raychaudhuri 
equation when studying the geometric properties 
of accelerating planes, we shall instead apply 
the linear field approximation of Einstein's 
field equation. 

    The starting point of the linear field 
approximation is to write the spacetime metric 
as:
\begin{equation}
ds^2 = (\eta_{\mu\nu} + 
h_{\mu\nu})\,dx^\mu\,dx^\nu.
\end{equation}
In this equation, $x^\mu$'s are the flat 
Minkowski coordinates, and 
$\eta_{\mu\nu}:=diag(1,-1,-1,-1)$ is the flat 
Minkowski metric. The field $h_{\mu\nu}$ 
represents a small deviation from the flat 
spacetime geometry. The linear field 
approximation is therefore particularly useful 
for the investigation of the small changes in the 
spacetime geometry. In what follows, we shall 
assume that at the spacetime point 
$P=(t,x,y,z)=(0,0,0,1/a)$, where our accelerating 
observer is at rest with respect to the 
radiation, the field $h_{\mu\nu}$ and its first 
derivatives vanish. In other words, we shall 
assume that at the point $P$ spacetime metric is 
that of flat spacetime. When we move away from 
the point $P$, spacetime metric deviates slightly 
from its flat spacetime form. We shall also 
assume that the spacetime metric (3.1) remains 
unchanged when we interchange the spatial 
coordinates representing deviations from the 
point P. In other words, we shall assume that, 
when looked from the point $P$, spacetime 
geometry looks the same in all spatial 
directions. This requirement is motivated by the 
fact that the components of the energy momentum 
stress tensor $T^{\mu\nu}$ remain unchanged 
under the interchanges of the spatial coordinates 
of spacetime. It is reasonable to expect that the 
same invariance property is possessed also by the 
spacetime metric itself.

   When $h_{\mu\nu}$ satisfies the so called 
Hilbert gauge condition
\begin{equation}
\partial_\mu(h^{\mu\nu} - 
\frac{1}{2}\eta^{\mu\nu}h) = 0,
\end{equation}
the linear field approximation of Einstein's 
field equation may be written as:
\begin{equation}
\partial_\lambda\partial^\lambda(h^{\mu\nu} - 
\frac{1}{2}\eta^{\mu\nu}h) = -16\pi T^{\mu\nu}.
\end{equation}
In these equations, $h:=\eta^{\mu\nu}h_{\mu\nu}$, 
and the components of $T^{\mu\nu}$ are given by 
Eqs.(2.1)-(2.3). Since the electromagnetic 
radiation which fills spacetime is assumed to be in thermal 
equilibrium, and therefore homogeneous, we shall 
look for a solution, where the diagonal elements 
of the metric depend on the time $t$ only. It is 
easy to see that there is just one solution to 
Eqs.(3.2) and (3.3), which satisfies this 
requirement, together with the other two 
requirements given before. The solution in 
question is the one, where the spacetime metric 
(3.1) takes the form:
\begin{eqnarray}
ds^2 &=& (1 - 8\pi\rho t^2)\,dt^2 - 
\frac{32}{3}\pi\rho t(x\,dx + y\,dy + 
\tilde{z}\,d\tilde{z})\,dt \nonumber \\ & &-(1 + 
\frac{8}{3}\pi\rho t^2)(dx^2 + dy^2 + 
d\tilde{z}^2) - \frac{16}{3}\pi\rho(xy\,dx\,dy + 
x\tilde{z}\,dx\,d\tilde{z} + 
y\tilde{z}\,dy\,d\tilde{z}),
\end{eqnarray}
where we have defined:
\begin{equation}
\tilde{z} := z - \frac{1}{a}.
\end{equation}

    We are now prepared to calculate the change 
in the area of a considered region of the 
accelerating $xy$-plane, from the point of view 
of an observer at rest with respect to that 
plane. In general, the area of a region on the 
$xy$-plane is
\begin{equation}
A = \int_{\mathcal{A}}\sqrt{q}\,dx\,dy,
\end{equation}
where $q$ is the determinant of the metric 
induced on that plane, and $\mathcal{A}$ is the 
domain of integration. Change in the area during 
the course of time is therefore a result from 
changes both in the metric and the domain of 
integration. When a plane is in accelerating 
motion in curved spacetime, the points of the 
region under consideration move on the plane with 
respect to the coordinates on that plane (or, 
rather, the coordinates move with respect to the 
points), and the geometry of the plane itself 
changes. Both of these contributions must be 
taken into account, when we calculate the area 
change. 

   To begin with, we investigate the movements of 
the points on the plane. Consider the point 
$Q=(x,y)$ on the $xy$-plane, which has the point
\begin{equation}
\mathcal{O} := (\frac{1}{a}\sinh(a\tau),0,0, 
\frac{1}{a}\cosh(a\tau))
\end{equation}
of spacetime as its origin. Our aim is to 
calculate the coordinates $(x',y')$ of $Q$ after 
the point $\mathcal{O}$ has been transported to the point 
$\mathcal{O}'$, where $\tau$ has been replaced by 
$\tau + d\tau$. We shall assume that when 
$\tau=0$, the point $Q$ is at rest with respect 
to the coordinates $x$, $y$ and $z$. In other 
words, we shall assume that when $\tau=0$, the 
only non-zero component of the four-velocity 
$u^\mu$ of the point $Q$ is
\begin{equation}
u^0 = 1.
\end{equation}
To calculate the four-velocity of $Q$, when $\tau 
\neq 0$, we parallel transport the vector $u^\mu$ 
from the point $(0,x,y,1/a)$ to the point 
$(\frac{1}{a}\sinh(a\tau),x,y, 
\frac{1}{a}\cosh(a\tau))$. In infinitesimal 
parallel transport from point $x^\mu$ to the 
point $x^\mu + dx^\mu$ the change experienced by 
$u^\mu$ is, in general,
\begin{equation}
\delta u^\mu = 
-\Gamma^\mu_{\alpha\beta}u^\alpha\,dx^\beta,
\end{equation}
where $\Gamma^\mu_{\alpha\beta}$ is the 
Christoffel symbol. In the linear field 
approximation $\Gamma^\mu_{\alpha\beta}$ takes 
the form:
\begin{equation}
\Gamma^\mu_{\alpha\beta} = 
\frac{1}{2}\eta^{\mu\sigma}(\partial_\beta 
h_{\sigma\alpha} + \partial_\alpha 
h_{\beta\sigma} - \partial_\sigma 
h_{\alpha\beta}).
\end{equation}
Using Eqs.(3.4) and (3.9) we find that at the 
point $(\frac{1}{a}\sinh(a\tau),x,y,
\frac{1}{a}\cosh(a\tau))$ we have:
\begin{subequations}
\begin{eqnarray}
u^1 &=& -\frac{16}{3}\pi\rho x\tau,\\
u^2 &=& -\frac{16}{3}\pi\rho y\tau,
\end{eqnarray}
\end{subequations}
where we have kept the terms linear in $\tau$ 
only. When $\tau$ is changed to $\tau + d\tau$, 
the coordinates $(x,y)$ of the point $Q$ on the 
$xy$-plane transform to
\begin{subequations}
\begin{eqnarray}
x' &=& x + u^1\,d\tau = (1 - 
\frac{16}{3}\pi\rho\tau\,d\tau)x,\\
y' &=& y + u^2\,d\tau = (1 - 
\frac{16}{3}\pi\rho\tau\,d\tau)y.
\end{eqnarray}
\end{subequations}
In other words, we have just re-scaled the 
coordinates $x$ and $y$ by the factor $(1 - 
\frac{16}{3}\pi\rho\tau\,d\tau)$.

     It only remains to calculate the change in 
the metric. The metric induced on the $xy$-plane 
is
\begin{equation}
dL^2 = (1 + \frac{8}{3}\pi\rho t^2)(dx^2 + dy^2) 
- \frac{16}{3}\pi\rho xy\,dx\,dy.
\end{equation}
The change experienced by $\sqrt{q}$ when $\tau$ 
is changed to $\tau + d\tau$ is therefore:
\begin{equation}
\delta\sqrt{q} = \Big(\frac{\partial}{\partial 
t}\sqrt{q}\Big)\frac{dt}{d\tau}\,d\tau = 
\frac{16}{3}\pi\rho\tau\, d\tau,
\end{equation}
where we have, again, kept the terms linear in 
$\tau$ only. We have also neglected the terms 
non-linear in $h_{\mu\nu}$. The relationship 
between $t$ and $\tau$ is given by Eq.(2.6a). 
Combining Eqs.(3.6), (3.12) and (3.14) we find 
that during the proper time interval $d\tau$ the 
area $A$ of the region under consideration 
becomes to
\begin{equation}
A' = (1 + \frac{16}{3}\pi\rho\tau\,d\tau)(1 - 
\frac{16}{3}\pi\rho\tau\,d\tau)^2A.
\end{equation}
Because $d\tau$ is infinitesimal, the change 
experienced by $A$ is therefore:
\begin{equation}
dA = -\frac{16}{3}\pi A\rho\tau\,d\tau.
\end{equation}
As one can see, the plane shrinks, when radiation 
flows through the plane to the negative 
$z$-direction.

\section{Area and Entropy}

        We denoted in Eq.(2.11) the amount of 
energy carried by radiation through the 
$xy$-plane by $\delta Q$ for a very good reason: 
Energy comes through the $xy$-plane as {\it 
heat}. According to the first law of 
thermodynamics the change in the heat of a system 
may be written as:
\begin{equation}
\delta Q = dE + p\,dV,
\end{equation}
where $dE$ is the (infinitesimal) change in the 
total energy, and $dV$ in the volume of the 
system. It is easy to see that both terms on the 
right hand side of Eq.(4.1) are present on the 
right hand side of Eq.(2.11): The increase in the 
total energy in the spatial region "behind" the 
accelerating plane from the point of view of the 
accelerating observer during an infinitesimal 
proper time interval $d\tau$ is
\begin{equation}
dE = \frac{1}{2}A\rho\sinh(2a\tau)\,d\tau,
\end{equation}
and because the increase in the three-volume $V$ 
of space "behind" the accelerating plane is, from 
the point of view of the accelerating observer, 
\begin{equation}
dV = A\sinh(a\tau)\,d\tau,
\end{equation}
we find, using Eq.(2.3), that the "work term" is
\begin{equation}
p\,dV = \frac{1}{6}A\rho\sinh(2a\tau)\,d\tau.
\end{equation}
When put together, the terms $dE$ and $p\,dV$ 
give the right hand side of Eq.(2.11).

   After having convinced ourselves that the 
right hand side of Eq.(2.11) really gives the 
heat transported through the accelerating plane 
from the point of view of the accelerating 
observer, we may now turn to the relationship 
between area and entropy. Because between the 
infinitesimal changes in the heat $Q$ and the 
entropy $S$ of a system there is a relationship:
\begin{equation}
\delta Q = T\,dS,
\end{equation}
where $T$ is the absolute temperature of the 
system, and because it follows from Eq.(2.11) 
that, for very small $\tau$:
\begin{equation}
\delta Q = \frac{4}{3}A\rho a\tau\,d\tau,
\end{equation}
we find, using Eq.(3.16), that the entropy $dS$ 
carried by radiation through the accelerating 
plane is related to the change $dA$ in the area 
of that plane such that:
\begin{equation}
T\,dS = -\frac{a}{4\pi}\,dA.
\end{equation}

      It only remains to fix the absolute 
temperature $T$ of the electromagnetic radiation. 
At this point we turn to the Unruh effect. 
According to that effect an accelerating observer 
experiences himself to be immersed in a heat bath of 
thermal particles. The temperature of this heat 
bath is the Unruh temperature \cite{viides}
\begin{equation}
T_U := \frac{a}{2\pi}.
\end{equation}
Suppose that when $\tau=0$, spacetime is, from 
the point of view of the accelerating observer, 
filled with electromagnetic radiation in thermal 
equilibrium at the Unruh temperature $T_U$. When 
$\tau=0$, there is no net flow of heat in any 
direction in the accelerating observer's frame of 
reference. After a very short elapsed proper time 
$\tau$, however, there is a net heat flow to the 
negative $z$-direction in the observer's frame of 
reference, and the amount of heat transported 
through the $xy$-plane during a proper time 
interval $d\tau$ is given by Eq.(4.6). The 
temperature of this heat flow is still the Unruh 
temperature $T_U$, and we may substitute $T_U$ 
for $T$ in Eq.(4.7). When we perform this 
substitution, we get:
\begin{equation}
dS = -\frac{1}{2}\,dA.
\end{equation}
In other words, the entropy carried by radiation 
through the accelerating plane is, in natural 
units, exactly one half of the decrease in the 
area of that plane. This is the final result of 
this paper, and it holds whenever the temperature 
of the radiation is equal to the Unruh 
temperature of an observer at rest with respect 
to the accelerating plane. It is remarkable that 
we have obtained a simple linear dependence 
between area and entropy even when our 
accelerating two-plane is not a part of any 
horizon of spacetime.

    \section{Concluding Remarks}

    In this paper we have considered an explicit 
example of a process, where radiation flows 
through a spacelike two-surface in such a way 
that the entropy carried through that two-surface 
is proportional to the decrease  in its area even 
when that two-surface is not a part of any 
horizon of spacetime. More precisely, we 
considered the flow of electromagnetic radiation 
through an accelerating two-plane, and we found 
that if the temperature of the radiation is the 
Unruh temperature of an observer at rest with 
respect to the accelerating plane, the entropy 
carried by radiation through that plane is, in 
natural units, exactly one half of the decrease 
in the area of that plane. This result bears a 
close resemblance to the well-known results 
concerning the radiation emitted by the black 
hole, cosmological, and Rindler horizons of 
spacetime. According to those results the entropy 
carried by radiation out from a horizon is, in 
natural units, exactly one quarter of the 
decrease in its area, whereas we found the same 
relationship -with an important exception that 
the constant of proportionality is not one 
quarter but one half- between the entropy flow 
and the area decrease when the two-surface is not 
a horizon but an arbitrary accelerating 
two-plane.

   The derivation of our result should be rather 
uncontroversial. Indeed, we have used just the 
fundamental results of thermodynamics and 
classical general relativity, together with the 
basic properties of Unruh radiation. We first 
assumed that spacetime is filled with 
electromagnetic radiation in thermal equilibrium, 
and then calculated the heat flow through a 
uniformly accelerating two-plane to the direction 
opposite to the direction of the motion of that 
two-plane. When radiation flows through the 
two-plane, the plane shrinks, and the decrease in 
its area may be calculated by means of the 
linear field approximation. Using the first law of 
thermodynamics one obtains the relationship 
between the flow of entropy and the decrease in 
area. Finally, if one substitutes for the 
temperature the Unruh temperature measured by an 
observer at rest with respect to the accelerating 
two-plane, one obtains the core result of this 
paper.

   Our result seems to support the view that 
entropy is possessed not only by horizons but, in 
addition, by {\it all} accelerating, spacelike 
two-planes of spacetime. When radiation flows 
through a two-plane, the plane shrinks, and the 
entropy of  the two-plane is converted to the 
entropy of radiation. If one accepts the view 
that entropy proportional to area is possessed by 
all accelerating two-planes, there is only a 
small step left to the idea that entropy is 
possessed by {\it all} spacelike two-surfaces of 
spacetime, no matter whether the two-surface 
under consideration is a part of any horizon or 
not.

   An enigmatic feature of our analysis is that 
it suggests that the entropy of an accelerating 
two-plane is, in natural units, not one quarter 
as one might expect, but {\it one half} of its 
area. An explanation of this curious result may 
lie in the fact that the radiation process of a 
horizon, and the radiation process considered in 
this paper are completely different: When a 
horizon radiates, its geometry interacts with the 
quanta of its radiation by its {\it one side} 
only. When radiation flows through a two-plane, 
however, the quanta of radiation interact with 
the geometry of spacetime on the {\it both sides} 
of the two-plane. So it is possible that 
radiation picks up entropy from the both sides of 
the two-plane. This may explain why the numerical 
value given by our analysis for the entropy of an 
accelerating two-plane is exactly twice the 
entropy of a spacetime horizon with the same 
area.

\begin{acknowledgements}

I thank Jorma Louko and Ari Peltola for their constructive criticism
during the preparation of this paper.

\end{acknowledgements}

  \end{document}